# Equivalence of total force (and torque) for two formulations of the Lorentz law

Masud Mansuripur, Armis R. Zakharian, and Jerome V. Moloney
College of Optical Sciences, The University of Arizona, Tucson, AZ 85721

**Abstract**. Two formulations of the Lorentz law of force in classical electrodynamics yield identical results for the total force (and total torque) of radiation on a solid object. The object may be surrounded by the free space or immersed in a transparent dielectric medium such as a liquid. We discuss the relation between these two formulations and extend the proof of their equivalence to the case of solid objects immersed in a transparent medium.

**Introduction**. Computation of the force of radiation on a given object, through evaluation of the electromagnetic field distribution according to Maxwell's equations, followed by a direct application of the Lorentz law of force, has been described in our previous publications.[1-5] In particular, for an isotropic, piecewise-homogeneous dielectric medium, the total force was shown to result from the force of the magnetic field acting on the induced bound current density, $\boldsymbol{J}_b \times \boldsymbol{B} = [(1 - 1/\varepsilon)\nabla \times \boldsymbol{H}] \times \boldsymbol{B}$, and from the force exerted by the electric component of the light field on the induced bound charge density at the interfaces between media of differing relative permittivity $\varepsilon$. The contribution of the $E$-field component of the Lorentz force is thus given by the force density $\rho_b \boldsymbol{E} = (\varepsilon_0 \nabla \cdot \boldsymbol{E})\boldsymbol{E}$. In a recent publication, Barnett and Loudon have shown that an alternative formulation of the Lorentz law – one that is often used in the radiation pressure literature – leads to exactly the same results as far as the total force and torque of radiation on a solid object are concerned.[6] We discuss the relation between these two formulations and extend the proof of their equivalence to the case of solid objects immersed in a transparent dielectric medium.

**Two Formulations of the Lorentz Law**. The Lorentz law of force, $\boldsymbol{F} = q(\boldsymbol{E} + \boldsymbol{V} \times \boldsymbol{B})$, may be written in two different ways for a medium in which the macroscopic Maxwell equations are satisfied. The two formulations are:

$$\boldsymbol{F}_1(\boldsymbol{r}) = -(\nabla \cdot \boldsymbol{P})\boldsymbol{E} + (\partial \boldsymbol{P}/\partial t) \times \boldsymbol{B} \qquad (1)$$

$$\boldsymbol{F}_2(\boldsymbol{r}) = (\boldsymbol{P} \cdot \nabla)\boldsymbol{E} + (\partial \boldsymbol{P}/\partial t) \times \boldsymbol{B} \qquad (2)$$

In an isotropic, homogeneous medium of dielectric constant $\varepsilon$, where the electric displacement field is $\boldsymbol{D}(\boldsymbol{r}) = \varepsilon_0 \boldsymbol{E}(\boldsymbol{r}) + \boldsymbol{P}(\boldsymbol{r}) = \varepsilon_0 \varepsilon \boldsymbol{E}(\boldsymbol{r})$, one may write $\boldsymbol{P}(\boldsymbol{r}) = \varepsilon_0(\varepsilon - 1)\boldsymbol{E}(\boldsymbol{r})$. In the absence of free charges and free currents in such a medium, Maxwell's first equation, $\nabla \cdot \boldsymbol{D}(\boldsymbol{r}) = 0$, implies that the volume density of bound charges within the medium is zero, that is, $\rho_b = -\nabla \cdot \boldsymbol{P}(\boldsymbol{r}) = 0$. This leaves for the Lorentz force density in the bulk of the medium, according to Eq. (1), only the magnetic contribution, namely, $\boldsymbol{F}_1(\boldsymbol{r}) = (\partial \boldsymbol{P}/\partial t) \times \boldsymbol{B}$. The formulation according to Eq. (2), however, leads to an entirely different result. Using Maxwell's equations in conjunction with the Lorentz law as expressed in Eq. (2) yields:

$$\boldsymbol{F}_2(\boldsymbol{r}) = \tfrac{1}{4}\,\varepsilon_0(\varepsilon - 1)\nabla(|\boldsymbol{E}_x|^2 + |\boldsymbol{E}_y|^2 + |\boldsymbol{E}_z|^2). \qquad (3)$$

In other words, the Lorentz force density in the bulk of a homogeneous medium according to the second formulation is proportional to the gradient of the *E*-field intensity, irrespective of the state of polarization of the optical field.

At the boundaries of isotropic media, the two formulations again lead to substantially different force densities. The contribution of the *B*-field to the surface force is negligible as, for non-magnetic media ($\mu = 1$), the *B*-field is continuous at the boundary; also, the discontinuity of $\partial \boldsymbol{P}/\partial t$, if any, is finite. In contrast, the perpendicular component of the *E*-field at the boundary, $\boldsymbol{E}_\perp$, has a sharp discontinuity, which results in a Dirac $\delta$-function behavior for the *E*-field gradient $\nabla E_\perp(\boldsymbol{r})$. When evaluating the integral of $(\boldsymbol{P}\cdot\nabla)\boldsymbol{E}$ across the boundary between the free space and a medium of dielectric constant $\varepsilon$, one finds a force density (per unit interfacial area) $\boldsymbol{F}_2(\boldsymbol{r}) = \tfrac{1}{2}P_\perp(\boldsymbol{E}_{a\perp} - \boldsymbol{E}_{b\perp})$, where $P_\perp$ is the normal component of the medium's polarization density at the surface, $\boldsymbol{E}_{a\perp}$ is the perpendicular *E*-field in the free space region just outside the medium, and $\boldsymbol{E}_{b\perp}$ is the perpendicular *E*-field within the medium just beneath the surface.[6] Note that $P_\perp$, being identical to the bound surface charge density $\sigma_b$, may be derived from the discontinuity of the magnitude of $\boldsymbol{E}_\perp$ at the surface, namely, $P_\perp = \sigma_b = \varepsilon_o(E_{a\perp} - E_{b\perp})$.

In the first formulation, the surface charge density $\sigma_b$ derived from $-(\nabla\cdot\boldsymbol{P})$ is equal to $P_\perp$, as mentioned before. This $\sigma_b$, however, must be multiplied by the effective *E*-field at the boundary to yield the surface force density (per unit area) $\boldsymbol{F}_1(\boldsymbol{r}) = \sigma_b\boldsymbol{E}$. Here the tangential component $\boldsymbol{E}_\parallel$ of the *E*-field is continuous at the boundary and, therefore, defined unambiguously. The perpendicular component, however, can be shown to be exactly equal to the average $\boldsymbol{E}_\perp$ at the boundary, namely, $\boldsymbol{E}_\perp = \tfrac{1}{2}(\boldsymbol{E}_{a\perp} + \boldsymbol{E}_{b\perp})$.

We give two examples from electrostatics to demonstrate the difference between the two formulations of the Lorentz law as concerns the *E*-field contribution to the force within isotropic media. In the first example, shown on the left-hand side of Fig. 1, the medium is heterogeneous with a dielectric constant $\varepsilon(x)$. Let the *D*-field be constant and oriented along the *x*-axis, that is, $\boldsymbol{D}(x) = D_o\hat{\boldsymbol{x}} = [\varepsilon_o E_x(x) + P_x(x)]\hat{\boldsymbol{x}} = \varepsilon_o\varepsilon(x)E_x(x)\hat{\boldsymbol{x}}$. (Clearly $\nabla\cdot\boldsymbol{D} = 0$ and $\nabla\times\boldsymbol{E} = 0$, as required by Maxwell's electrostatic equations.) From the first formulation, Eq. (1), we find:

$$F_{1x}^{surface}(x=0) = -\tfrac{1}{2}P_x(0)\,[E_o + E_x(0)] = -(D_o/2\varepsilon_o)[1 + 1/\varepsilon(0)]P_x(0); \qquad x = 0 \tag{4}$$

$$F_{1x}^{bulk}(x) = -E_x(x)\,\mathrm{d}P_x(x)/\mathrm{d}x = (1/2\varepsilon_o)\mathrm{d}[D_o - P_x(x)]^2/\mathrm{d}x; \qquad 0 < x < L \tag{5}$$

$$F_{1x}^{surface}(x=L) = (D_o/2\varepsilon_o)[1 + 1/\varepsilon(L)]P_x(L); \qquad x = L. \tag{6}$$

The second formulation, Eq. (2), yields:

$$F_{2x}^{surface}(x=0) = \tfrac{1}{2}P_x(0)\,[E_x(0) - E_o] = -(D_o/2\varepsilon_o)[1 - 1/\varepsilon(0)]P_x(0); \qquad x = 0 \tag{7}$$

$$F_{2x}^{bulk}(x) = P_x(x)\,\mathrm{d}E_x(x)/\mathrm{d}x = -(1/2\varepsilon_o)\mathrm{d}P_x^2(x)/\mathrm{d}x; \qquad 0 < x < L \tag{8}$$

$$F_{2x}^{surface}(x=L) = (D_o/2\varepsilon_o)[1 - 1/\varepsilon(L)]P_x(L); \qquad x = L. \tag{9}$$

Clearly the two formulations predict different distributions for the force density (both surface and bulk). However, the total force – obtained by integrating the bulk density and adding the surface contributions – turns out to be exactly the same in the two formulations ($\boldsymbol{F}^{total} = 0$).

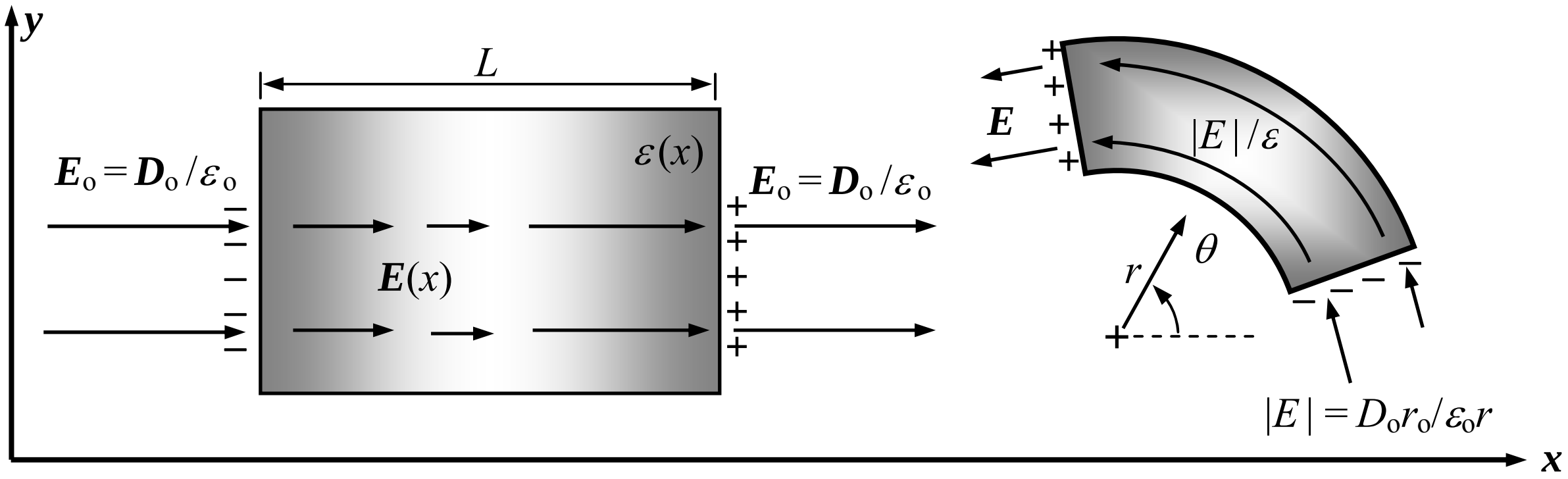

**Figure 1**. Examples of electrostatic field distributions within dielectric media. On the left-hand side, the medium's dielectric constant $\varepsilon(x)$ varies along the $x$-axis while the electric displacement $\boldsymbol{D} = D_o\hat{\boldsymbol{x}}$ remains constant. On the right-hand side, $\varepsilon$ is constant within the crescent-shaped medium ($r_o < r < r_1$, $\theta_0 < \theta < \theta_1$), while $\boldsymbol{D}(\boldsymbol{r}) = (D_o r_o/r)\hat{\boldsymbol{\theta}}$, which is azimuthally oriented, decreases with the inverse radius $r$.

In the second example, shown on the right-hand-side of Fig. 1, the medium is homogeneous (dielectric constant $= \varepsilon$), and the $D$-field profile is $\boldsymbol{D}(\boldsymbol{r}) = (D_o r_o/r)\hat{\boldsymbol{\theta}}$. In the first formulation, based on Eq. (1), the bulk force is zero, and the surface force density is

$$\boldsymbol{F}_1^{surface}(r, \theta = \theta_{0,1}) = \pm(1/2\varepsilon_o)(1 - 1/\varepsilon^2)\,(D_o r_o/r)^2\hat{\boldsymbol{\theta}}; \qquad \theta = \theta_0, \theta_1 \tag{10}$$

In the second formulation, Eq. (2) yields:

$$\boldsymbol{F}_2^{surface}(r, \theta = \theta_{0,1}) = \pm(1/2\varepsilon_o)(1 - 1/\varepsilon)^2(D_o r_o/r)^2\hat{\boldsymbol{\theta}}; \qquad \theta = \theta_0, \theta_1 \tag{11}$$

$$\boldsymbol{F}_2^{bulk}(r, \theta) = -(1/\varepsilon_o\varepsilon)(1 - 1/\varepsilon)(D_o^2 r_o^2/r^3)\,\hat{\boldsymbol{r}}; \qquad \theta_0 < \theta < \theta_1 \tag{12}$$

Once again, the force distributions in the two formulations are quite different, but the total force is exactly the same, that is,

$$\boldsymbol{F}^{total} = -(1/\varepsilon_o)(1 - 1/\varepsilon^2)\,(D_o r_o/r)^2 \sin[\tfrac{1}{2}(\theta_1 - \theta_0)]\,\hat{\boldsymbol{r}}. \tag{13}$$

Next we prove the generality of the above results by showing that the total force obtained from Eqs. (1) and (2) is the same under all circumstances. The proof of equivalence between the two formulations with regard to total force (and also total torque) is due to Barnett and Loudon,[6] who also clarified the role of surface forces in the second formulation. In what follows we outline this proof of equivalence along the same lines as originally suggested by Barnett and Loudon, then extend their results to objects that are immersed in a liquid (or surrounded by an isotropic, homogeneous medium of a differing index of refraction).

Consider the difference between $\boldsymbol{F}_1(\boldsymbol{r})$ and $\boldsymbol{F}_2(\boldsymbol{r})$, written as

$$\Delta\boldsymbol{F}(\boldsymbol{r}) = \boldsymbol{F}_2(\boldsymbol{r}) - \boldsymbol{F}_1(\boldsymbol{r}) = (\boldsymbol{P}\cdot\nabla)\boldsymbol{E} + (\nabla\cdot\boldsymbol{P})\boldsymbol{E} = \partial(P_x\boldsymbol{E})/\partial x + \partial(P_y\boldsymbol{E})/\partial y + \partial(P_z\boldsymbol{E})/\partial z. \tag{14}$$

When $\int \Delta \boldsymbol{F}(\boldsymbol{r})\,\mathrm{d}x\mathrm{d}y\mathrm{d}z$ over the volume of an object is evaluated, individual terms on the right-hand-side of Eq. (14), being complete differentials, produce $(P_x \boldsymbol{E})|_{x_\max} - (P_x \boldsymbol{E})|_{x_\min}$, etc., after a single integration over the proper variable. These expressions then reduce to zero because $x_{\min}, x_{\max}$ are either in the free-space region outside the object, where $\boldsymbol{P} = 0$, or outside the beam's boundary, where $\boldsymbol{E} = \boldsymbol{P} = 0$. Either way, the volume integral of $\Delta \boldsymbol{F}(\boldsymbol{r})$ evaluates to zero, yielding $\int \boldsymbol{F}_1(\boldsymbol{r})\,\mathrm{d}x\mathrm{d}y\mathrm{d}z = \int \boldsymbol{F}_2(\boldsymbol{r})\,\mathrm{d}x\mathrm{d}y\mathrm{d}z$.

If the object of interest happens to be immersed in a liquid (or surrounded by a medium other than the free space), where, for example, the condition $(P_x \boldsymbol{E})|_{x_\max} = 0$ cannot be ascertained, one must proceed to assume the existence of a narrow gap between the object and its surroundings, as shown in Fig. 2. Under such circumstances, the radiation force exerted on the boundary of the object should be computed using $\boldsymbol{E}_g(\boldsymbol{r})$, the gap field, rather than $\boldsymbol{E}_a(\boldsymbol{r})$, the field in the immersion medium just outside the object.[2] However, the introduction of the gap creates oppositely charged layers (across the gap), and the force of attraction between these proximate charges accounts for the difference between $\boldsymbol{F}^{surface}(\boldsymbol{r})$ computed using $\boldsymbol{E}_g(\boldsymbol{r})$ and $\boldsymbol{E}_b(\boldsymbol{r})$ on the one hand, and that computed using $\boldsymbol{E}_a(\boldsymbol{r})$ and $\boldsymbol{E}_b(\boldsymbol{r})$ on the other hand. Once the effect of this attractive force on the object is discounted, the remaining force may be computed by ignoring the gap field and assuming, for the first formulation, that the effective $E$-field is $\tfrac{1}{2}(\boldsymbol{E}_a + \boldsymbol{E}_b)$, while, for the second formulation, the $E$-field discontinuity is $\boldsymbol{E}_a - \boldsymbol{E}_b$.

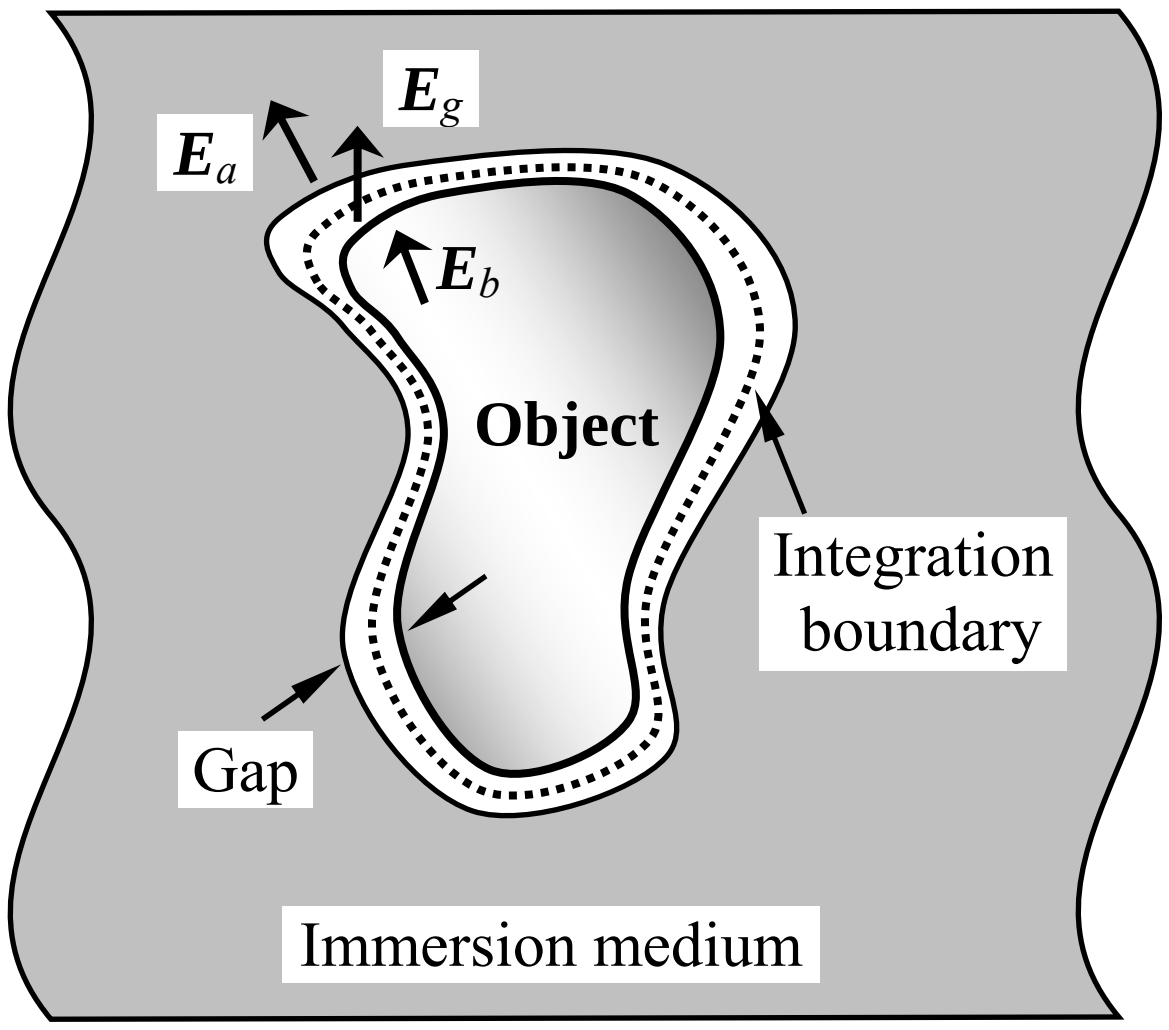

**Figure 2**. When the object is surrounded by a medium other than the free space, a narrow gap may be imagined to exist between the object and its surroundings, so that, within the gap, the polarization density $\boldsymbol{P}(\boldsymbol{r})$ may be set to zero. The integration boundary is then placed within the gap to ensure that the integrated force densities corresponding to Eqs.(1) and (2) lead to the same total force.

Finally, we show that the total torque $\boldsymbol{T}^{total}$ exerted on a given object is independent of whether $\boldsymbol{F}_1(\boldsymbol{r})$ or $\boldsymbol{F}_2(\boldsymbol{r})$ is used to compute the torque. In the following derivation, the fact that $\boldsymbol{P}(\boldsymbol{r}) = 0$ when $\boldsymbol{r}$ is outside the proper boundaries of the object is used in both steps. In the first step, terms containing the integrals of complete differentials are omitted. (As before, such integrals reduce to expressions that vanish outside the object's boundaries.) In the second step, integration by parts is used to simplify the integrands.

$$\Delta \boldsymbol{T}^{total} = \iiint \boldsymbol{r} \times \Delta \boldsymbol{F}(\boldsymbol{r})\, \mathrm{d}x\mathrm{d}y\mathrm{d}z = \iiint \boldsymbol{r} \times [\partial(P_x \boldsymbol{E})/\partial x + \partial(P_y \boldsymbol{E})/\partial y + \partial(P_z \boldsymbol{E})/\partial z]\, \mathrm{d}x\mathrm{d}y\mathrm{d}z$$

$$= \iiint \{ [y\partial(P_y E_z)/\partial y - z\partial(P_z E_y)/\partial z]\,\hat{\boldsymbol{x}} + [z\partial(P_z E_x)/\partial z - x\partial(P_x E_z)/\partial x]\,\hat{\boldsymbol{y}}$$

$$+ [x\partial(P_x E_y)/\partial x - y\partial(P_y E_x)/\partial y]\,\hat{\boldsymbol{z}} \}\, \mathrm{d}x\mathrm{d}y\mathrm{d}z$$

$$= - \iiint [(P_y E_z - P_z E_y)\,\hat{\boldsymbol{x}} + (P_z E_x - P_x E_z)\,\hat{\boldsymbol{y}} + (P_x E_y - P_y E_x)\,\hat{\boldsymbol{z}}]\mathrm{d}x\mathrm{d}y\mathrm{d}z = \iiint (\boldsymbol{E} \times \boldsymbol{P})\mathrm{d}x\mathrm{d}y\mathrm{d}z. \quad (15)$$

In an isotropic medium $\boldsymbol{E}$ and $\boldsymbol{P}$ will be parallel to each other and, therefore, their cross-product appearing on the right-hand-side of Eq. (15) will be zero. Consequently $\Delta \boldsymbol{T}^{total} = 0$, completing the proof that both formulations of the Lorentz law yield the same overall torque on the object.

**Concluding Remarks**. It is remarkable that the two formulations in Eqs. (1) and (2) yield the same total force (and torque) exerted on a given object under quite general circumstances. In fact, until recently, the present authors had relied solely on the first formulation, assuming (falsely) that the second formulation is incapable of producing consistent answers for certain well-defined problems.[1-5] This situation has now changed with the equivalence proof provided by Barnett and Loudon.[6] It appears, therefore, that the classical electromagnetic theory (with its reliance on the macroscopic properties of matter, such as polarization density $\boldsymbol{P}$, displacement field $\boldsymbol{D}$, dielectric constant $\varepsilon$, etc.) cannot decide which formula, if any, provides the correct force *distribution* throughout the object (even as both formulas yield the correct values for total force and torque). A resolution of this interesting problem may have to await the verdict of future experiments.

**Acknowledgments**. The authors are grateful to Rodney Loudon and Stephen Barnett for illuminating discussions and for sharing the preprint of their latest manuscript, Ref. [6], with us. This work has been supported by the Air Force Office of Scientific Research (AFOSR) contracts FA9550-04-1-0213, FA9550-04-1-0355, and F49620-03-1-0194.

1. M. Mansuripur, "Radiation pressure and the linear momentum of the electromagnetic field," *Optics Express* **12**, 5375-5401 (2004), http://www.opticsexpress.org/abstract.cfm?id=81636.
2. M. Mansuripur, A. R. Zakharian, and J. V. Moloney, "Radiation pressure on a dielectric wedge," *Optics Express* **13**, 2064-2074 (2005), http://www.opticsexpress.org/abstract.cfm?id=83011.
3. M. Mansuripur, "Radiation pressure and the linear momentum of light in dispersive dielectric media," *Optics Express* **13**, 2245-2250 (2005), http://www.opticsexpress.org/abstract.cfm?id=83032.
4. M. Mansuripur, "Angular momentum of circularly polarized light in dielectric media," *Optics Express* **13**, 5315-5324 (2005), http://www.opticsexpress.org/abstract.cfm?id=84895.
5. M. Mansuripur, "Radiation pressure and the distribution of electromagnetic force in dielectric media," *SPIE Proc.* **5930**, *Optical Trapping and Optical Micromanipulation II*, K. Dholakia and G. C. Spalding, Eds. (2005).
6. S. M. Barnett and R. Loudon, "On the electromagnetic force on a dielectric medium," *J. Phys. B: At. Mol. Opt. Phys.* **39**, S671-S684 (2006).